\documentclass[twocolumn,showpacs,amsmatsh,amssymb,aps,prc,groupedaddress,superscriptaddress]{revtex4-1}
\usepackage[caption=false]{subfig}
\usepackage{tabularx}
\usepackage{graphicx}% Include figure files
\usepackage{dcolumn}% Align table columns on decimal point
\usepackage{bm}% bold math
\usepackage{float}
\usepackage[Gray,squaren,thinqspace,thinspace]{SIunits} % Elegant eenheden zetten
\usepackage{amsmath}
\usepackage{ifthen}
\usepackage{cleveref}

\begin{document}

  %\preprint{}

\title{Lepton kinematics in low energy neutrino--Argon interactions}


\author{N.~Van Dessel}
\email{Nils.VanDessel@UGent.be}
\affiliation{Department of Physics and Astronomy,\\
Ghent University,\\ Proeftuinstraat 86,\\ B-9000 Gent, Belgium}
\author{A.~Nikolakopoulos}
\affiliation{Department of Physics and Astronomy,\\
Ghent University,\\ Proeftuinstraat 86,\\ B-9000 Gent, Belgium}
\author{N.~Jachowicz}
\email{Natalie.Jachowicz@UGent.be}
\affiliation{Department of Physics and Astronomy,\\
Ghent University,\\ Proeftuinstraat 86,\\ B-9000 Gent, Belgium}

  %\thanks{}


\date{\today}

\begin{abstract}
  \begin{description} 
\item[Background] Neutrinos in the low--energy regime provide a gateway to a wealth of interesting physics. While plenty of literature exists on detailing the calculation and measurement of total reaction strengths, relatively little attention is paid to the measurement and modeling of the final lepton through differential cross sections at low energies, despite the experimental importance.

\item[Purpose] We calculate differential cross sections for low--energy neutrino--nucleus scattering. We examine the role played by forbidden transitions in these distributions and how this differs across different energies and nuclear target masses. Attention is also paid to predictions for typical experimental neutrino spectra.

\item[Method] The differential cross sections are calculated within a Continuum Random Phase Approximation framework, which allows us to include collective excitations induced by long--range correlations. The Coulomb interaction of the final lepton in charged current events is treated in an effective way. 

\item[Results] Kinematic distributions are calculated for $^{16}$O, $^{40}$Ar and $^{208}$Pb. $^{40}$Ar model results are compared for CC ($\nu_e,e^-$) reactions to events generated by the MARLEY event generator~\cite{Gardiner2017}, with noticeable discrepancies.
 
\item[Conclusion] Forbidden transitions have a marked effect on the kinematic distributions of the final lepton at low--energy kinematics, such as for DAR neutrinos or for a Fermi--Dirac spectrum at low temperature. This could introduce biases in experimental analyses. Backwards scattering is noticeably more prominent than with MARLEY.
  \end{description}
\end{abstract}

\maketitle

\section{Introduction}\label{sec:int}
In recent times, neutrino physics has provided an exciting and rich area of research, with plenty of open questions either partially or completely unanswered. Major examples of these include the absolute mass hierarchy, the CP--violating phase and the possible existence of a fourth 'sterile' neutrino. Besides these fundamental high--energy physics issues, neutrinos are also important in other areas, such as e.g. cosmology, where the  mass of the neutrinos could have an effect on the expansion of our universe~\cite{Nagirner:2019igg}. \\
Of particular note is the role that neutrinos play in the realm of astrophysics. Here, the existence of massive neutrinos would e.g.~have an effect on galaxy formation~\cite{Fukugita:1984yk}. They're also an important part of supernovae, the explosive end to a sufficiently massive star's life cycle. In this process, neutrinos are produced in copious quantities in various flavors: electron neutrinos through the electron capture during core collapse and subsequent 'burst' as well as pair--produced neutrinos during the cool--down of the remnant protoneutron star. The energy carried by these neutrinos represents the biggest part of the star's gravitational binding energy, with the energy spectrum of the neutrinos being in the 10s of MeV range. An exact modeling of supernovae is highly dependent on the properties mentioned above, with the outgoing neutrino spectra depending both on the absolute mass hierarchy and oscillations with the Mikheyev--Smirnov--Wolfenstein (MSW) effect inside the exploding star~\cite{Ankowski:2016lab}. Interactions with nuclei, both charged--current and neutral current, will also influence these phenomenae, as well as have an effect on the nucleosynthesis that takes place in the supernovae envelope~\cite{Balasi:2015dba}. \\
Experimental efforts have been and will be undertaken to detect supernova neutrinos in the past, present and future. The Deep Underground Neutrino Experiment (DUNE), as well as e.g.~JUNO~\cite{juno} and Hyper-Kamiokande~\cite{hyperk} aim to make high--precision studies. The former will have the capacity to distinguish between the two possible neutrino mass hierarchies through detection of supernova neutrinos~\cite{Ankowski:2016lab}, as part of its low-energy program. Furthermore, through these signals, experiments could also very well unveil Beyond--Standard Model (BSM) physics, such as e.g. the aforementioned sterile neutrinos. DUNE will make use of a Liquid Argon Time Projection Chamber (LArTPC) as detector. Since supernovae are not available on demand, a more readily available source of low--energy neutrinos is required. This is e.g. possible at the Spallation Neutron Source (SNS) at Oakridge National Laboratory (ORNL)~\cite{Efremenko:2008an}, which provides an more readily available source of neutrinos of similar energies created out of pions decaying at rest, to perform measurements. \\
All of the experiments mentioned, as well as those performed in the past such as LSND and KARMEN~\cite{Auerbach:2001hz,Maschuw:1998qh}, have the neutrinos scattering off atomic nuclei ($^{40}$Ar for DUNE). It is therefore paramount that efforts are undertaken to provide an adequate theoretical modeling of the cross sections describing these processes for both NC and CC events. Not only are they crucial in the analyses of these experiments, but they are also needed to model the interaction of outgoing neutrinos in supernovae with the nuclei in the star and the subsequent nucleosynthesis. This is not trivial, as the nuclear response to low--energy neutrinos is highly dependent on the details of nuclear structure, and the description of the excitated states. While theoretical literature~\cite{McLaughlin:2004va,Engel:2002hg,Volpe:2001gy,Auerbach:1997ay,SUZUKI2003446,Bandyopadhyay:2016gkv,Cheoun:2011zza,Hayes:1999ew,Kostensalo:2018kgh,Kolbe:2000np,Kolbe:1999vc,Paar:2007fi,Samana:2008pt,Nieves:2004wx,SajjadAthar:2005ke,Singh:1998md,Kolbe:1999au,Jachowicz:2002hz,Volpe:2000zn,Pourkaviani:1990et,Fukugita:1988hg,Paar:2011pz,Suzuki:2009zzc,Cheoun:2010pn,GilBotella:2003sz,Kolbe:2003ys,Suzuki:2012ds,Suzuki:2006qd,Suzuki:2013wda} pertaining to low--energy neutrino interactions with nuclei is rich with detail on the calculations of total cross sections, the amount of attention paid to the description of the final lepton's kinematics, is comparatively modest. This stands in contrast with research performed in the medium energy range (few 100 MeV to a few GeV), where lepton kinematics are a key ingredient in the analyses: they are needed in the energy reconstruction process~\cite{Nikolakopoulos:2018sbo}.  Furthermore, inclusive differential cross sections only require the detection of the charged lepton, and can provide a powerful tool with which to scrutinize theoretical models in the low--energy regime. We will therefore focus this paper on differential cross sections which contain information on the outgoing lepton's kinematics, such as scattering angle, for low--energy CC and NC neutrinos in a CRPA approach, with a focus on ${}^{40}$Ar.

\section{Model}\label{sec:for}
We now cover the theoretical ingredients employed in our calculations. The cross section for electroweak scattering of neutrinos off atomic nuclei in the Giant Resonance (GR) and quasielastic (QE) regime is given by the following expression:

\begin{figure}
  \centering
  \includegraphics[width=0.85\columnwidth]{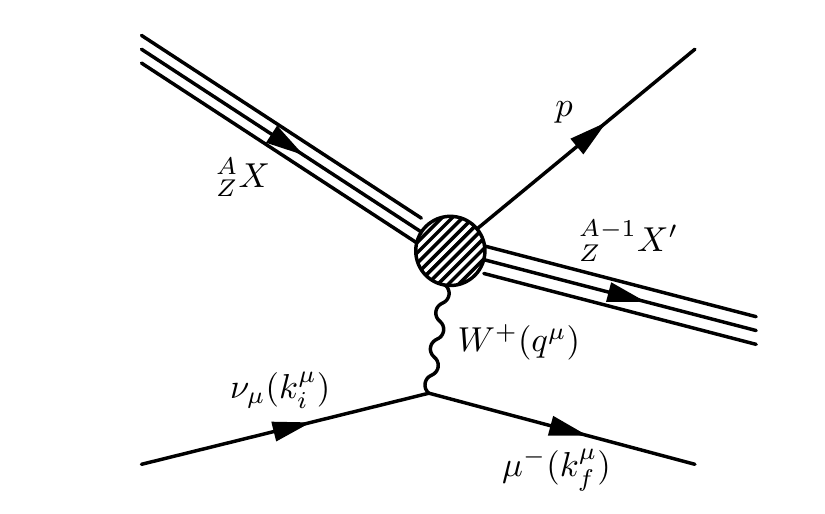}
  \caption{Diagrammatical representation of neutrino--nucleus scattering, pictured here for the case of a CC interaction.}
  \label{fig:diagram}
\end{figure}

\begin{equation}\label{eq:xsec}
\begin{aligned}
\frac{\mathrm{d}\sigma}{\mathrm{d}T_f\mathrm{d}\Omega_f} =& \sigma_X E_f k_f \zeta^2(Z',E_f) \\
&\times \left( v_{CC} W_{CC} + v_{CL} W_{CL} + v_{LL} W_{LL} \right.  \\
& + \left. v_{T} W_{T} \pm v_{T'} W_{T'} \right),
\end{aligned}
\end{equation}
differential in $T_f$ and $\Omega_f = (\theta_f,\phi_f)$, the kinetic energy and scattering angle of the outgoing lepton. It is furthermore a function of $E_f$ and $k_f$, the energy and momentum of the lepton. The Mott--like prefactor $\sigma_X$ is $\left(\frac{G_F \cos{\theta_c}}{2\pi} \right)^2$ for CC interactions scattering and $\left(\frac{G_F}{2\pi} \right)^2$ in the case of NC interactions. $G_F$ is the Fermi constant, which encodes the strength of the weak interaction, with $\cos{\theta_c}$ the cosine of the Cabibbo angle. The factor $\zeta^2(Z',E_f)$ accounts for the Coulomb interaction between the escaping lepton and the residual nucleus in case the reaction is CC, which we will come back to shortly. The $\pm$--sign differs between the case of neutrino and antineutrino as a result of the parity--breaking nature of the weak interaction, which depends on the neutrino's helicity. The $v$--factors are purely a function of the leptonic kinematics:
\begin{equation}\label{eq:inclusivevlist}
\begin{aligned}
v_{CC} &= 1+\beta \cos{\theta_f} ,\\
v_{CL} &= -\left( \frac{\omega}{q}\left( 1+\beta \cos{\theta_f} \right) + \frac{m_f^2}{E_fq} \right) ,\\
v_{LL} &= 1+\beta \cos{\theta_f} - \frac{2 E_i E_f}{q^2} \beta^2 \sin^2{\theta_f} ,\\
v_{T} &= 1-\beta \cos{\theta_f} + \frac{E_i E_f}{q^2} \beta^2 \sin^2{\theta_f} ,\\
v_{T'} &= \frac{E_i + E_f}{q}\left( 1-\beta \cos{\theta_f}  \right) - \frac{m_f^2}{E_fq} ,
\end{aligned}
\end{equation}
with $\omega$, $q$, $E_i$ and $m_f$ the energy transfer, momentum transfer, incoming neutrino energy and outgoing lepton mass, respectively and $\beta = \frac{k_f}{E_f}$. The $W$--factors are the nuclear response functions, which are dependent on the transition amplitudes between the initial ($| \Phi_\textrm{0} \rangle$) and final ($| \Phi_\textrm{f} \rangle$) state, and contain all the nuclear information involved in this process:

\begin{equation}\label{eq:inclusivewlist}
\begin{aligned}
W_{CC} &=  \sum_{J \geq 0} \sum_{l,j,j_h} \left| \langle \Psi_C^{(+)} (J) || \widehat{\mathcal{M}}^{C}_{J}(q) || \Phi_0 \rangle \right|^2,\\
W_{CL} &=  -2 \sum_{J \geq 0} \sum_{l,j,j_h} \mathrm{Re} \left[ \langle \Psi_C^{(+)} (J) || \widehat{\mathcal{M}}^{C}_{J}(q) || \Phi_0 \rangle \right. \\
&\times \left. \left(\langle \Psi_C^{(+)} (J) || \widehat{\mathcal{L}}^{L}_{J}(q) || \Phi_0 \rangle \right)^* \right],\\
W_{LL} &=  \sum_{J \geq 0} \sum_{l,j,j_h} \left| \langle \Psi_C^{(+)} (J) || \widehat{\mathcal{L}}^{L}_{J}(q) || \Phi_0 \rangle \right|^2,\\
W_{T} &=  \sum_{J \geq 1} \sum_{l,j,j_h} \left( \left| \langle \Psi_C^{(+)} (J) || \widehat{\mathcal{T}}^{E}_{J}(q) || \Phi_0 \rangle \right|^2 \right. \\
&+ \left. \left| \langle \Psi_C^{(+)} (J) || \widehat{\mathcal{T}}^{M}_{J}(q) || \Phi_0 \rangle \right|^2 \right),\\
W_{T'} &=  2 \sum_{J \geq 1} \sum_{l,j,j_h} \mathrm{Re} \left[ \langle \Psi_C^{(+)} (J) || \widehat{\mathcal{T}}^{E}_{J}(q) || \Phi_0 \rangle \right. \\
&\times \left.\left(\langle \Psi_C^{(+)} (J) || \widehat{\mathcal{T}}^{M}_{J}(q) || \Phi_0 \rangle \right)^* \right],
\end{aligned}
\end{equation}
with the angular momentum labels $J, j, l$ and $j_h$ refering to the multipole moment of the operator, the total and spatial angular momentum of the outgoing nucleon, and the total angular momentum of the remnant nucleus, respectively. The expressions for the Coulomb ($\hat{\mathcal{M}}_{J,L}^{C}$), longitudinal ($\hat{\mathcal{L}}_{J,L}^{L}$), electric ($\hat{\mathcal{T}}_{J,L}^{E}$) and magnetic ($\hat{\mathcal{T}}_{J,L}^{M}$) multipole operators similarly available in ~\cite{walecka2004theoretical}. For sufficiently low--energy neutrinos, one can employ the 'allowed approximation' (AA): if one assumes that the value of $q$ is negligible in the transition amplitudes (ergo, one uses the long--wavelength limit $q \rightarrow 0$) and one also assumes that one is dealing with slow nucleons $p_N/m_N \rightarrow 0$, it can be shown that the only surviving terms in the above multipole decomposition of the CC nuclear current (similar considerations hold for the NC current) are the following:
\begin{equation}
\begin{aligned}
\hat{\mathcal{M}}_{0,0}^{C} =& \frac{1}{\sqrt{4}}F_1 \tau_{\pm}(i)\\
\hat{\mathcal{T}}_{1,m}^{E} =& \sqrt{2} \hat{L}_{1,m}^{L} = \frac{i}{\sqrt{6}}G_A\sum_{i=1}^{A}\tau_{\pm}(i)\sigma_{1,m}(i),
\end{aligned}
\end{equation}
where $F_1$ and $G_A$ are the Fermi and axial form factors, respectively. These operators are the well--known Fermi and Gamow--Teller transition operators. The AA is an adequate description of the nucleus' response to an electroweak probe at low energies. Because they are responsible for the largest part of the reaction strength in this kinematic regime, they give rise to the 'allowed' transitions with the well--known selection rules through operators with $J^P$ quantum numbers of $0^+$ and $1^+$, respectively. Higher--order transitions, to draw a contrast, are often refered to as 'forbidden' transitions. In this work, the responses are calculated in the Continuum Random Phase Approximation (CRPA), where long--range correlations and collective excitations of the nucleus are taken into account. This scheme has seen succesful applications in the past. The details of how this approach can be used to calculate the nuclear response functions can be found in Refs.~\cite{Ryckebusch:1988aa, Ryckebusch:1989nn, Jachowicz:1998fn, Jachowicz:2002rr,Jachowicz:2002hz,Jachowicz:2004we, Jachowicz:2006xx, Jachowicz:2008kx,Pandey:2014tza, Pandey:2016jju,VanDessel:2017ery,VanDessel:2019atx}. Keeping in line with these previous works, we make use of the free-nucleon value for the axial coupling of $g_A = 1.27$. Some models use the 'quenched' $g_A = 1.00$ as detailed in Ref.~\cite{Pastore:2017uwc}, where it is shown that this effective value is needed due to the model space being truncated as well as not fully taking into account the effects of nuclear correlations. \\

At low energies, for CC interactions, the Coulomb attraction or repulsion between the residual nucleus and the outgoing lepton has a large effect on the cross section and needs to be properly accounted for. In principle, this can be achieved by considering the asymptotic lepton wave function as a sum of distorted partial waves, calculated in a Coulomb potential. If one only takes the S--wave into account (valid at low outgoing lepton momentum $p_f$), the ratio between the distorted and undistorted S--wave leads to the Fermi function~\cite{Engel:1997fy}:

\begin{equation}
\begin{aligned}
\zeta(Z',E_f)^2 =& 2(1+\gamma_0)(2p_f R)^{-2(1-\gamma_0)} \frac{|\Gamma(\gamma_0+i\eta)|^2}{(\Gamma(2\gamma_0+1))^2}e^{\pi\eta},
\end{aligned}
\end{equation}
with $R\approx 1.2 A^{1/3} \mathrm{fm}$ the nuclear radius, $\gamma_0=\sqrt{1-(\alpha Z')^2}$, $E_f$ the outgoing lepton's energy, $p_f$ the outgoing momentum and $\eta=\pm \frac{\alpha Z' c}{v}$. with $+$ and $-$ for neutrinos and antineutrinos respectively. The residual nucleus' electric charge $Z'$ is equal to $Z+1$ or $Z-1$ for $\nu$/$\bar{\nu}$, respectively. This approximation is not applicable once the lepton's outgoing momentum becomes appreciably high~\cite{Engel:1997fy}. The modified effective momentum approximation, detailed in Ref.~\cite{Engel:1997fy}, can be used in such a regime. This semi--classical approach consists of shifting the energy and momentum of the final lepton to an effective value by the Coulomb energy in the center of the nucleus:
\begin{equation}
E_{eff} = E_f - V_c(0) = E \pm \frac{3}{2}\frac{Z' \alpha \hbar c }{R}.
\end{equation}
This also introduces a factor in the differential cross section as a result of a change in the available phase space for the final lepton:
\begin{equation}
\zeta(Z',E_f)^2 = \frac{E_{eff} k_{eff} }{E_f k_f},
\end{equation}
and furthermore requires a shift in the momentum transfer $q \rightarrow q_{eff}$ in the the amplitudes in Eq.~\ref{eq:inclusivewlist}. In practice, we will interpolate between these two schemes that consists of taking the value of $\zeta(Z',E_f)^2$ that is closest to unity. This corresponds to taking the Fermi function for low $p_f$, and using MEMA at higher $p_f$. \\

The framework described above has the attractive property of providing a model that can be employed µfor a broad range of energies. It is capable of describing the Giant Resonance region, dominated by collective excitations, but can also (as discussed in Refs.~\cite{Pandey:2014tza, Pandey:2016jju,VanDessel:2017ery}) describe the quasielastic peak for higher energy regimes such as those seen in experiments such as T2K, MiniBooNE and MicroBooNE. Previous work on the topic of the interactions low--energy neutrino with nuclei such as $^{40}$Ar in the CRPA framework, including comparison with other models and Gamow--Teller strengths can be found in Ref.~\cite{VanDessel:2019atx}.

\section{Results}\label{sec:res}

\begin{figure}
   \centering
   \includegraphics[width=0.95\columnwidth]{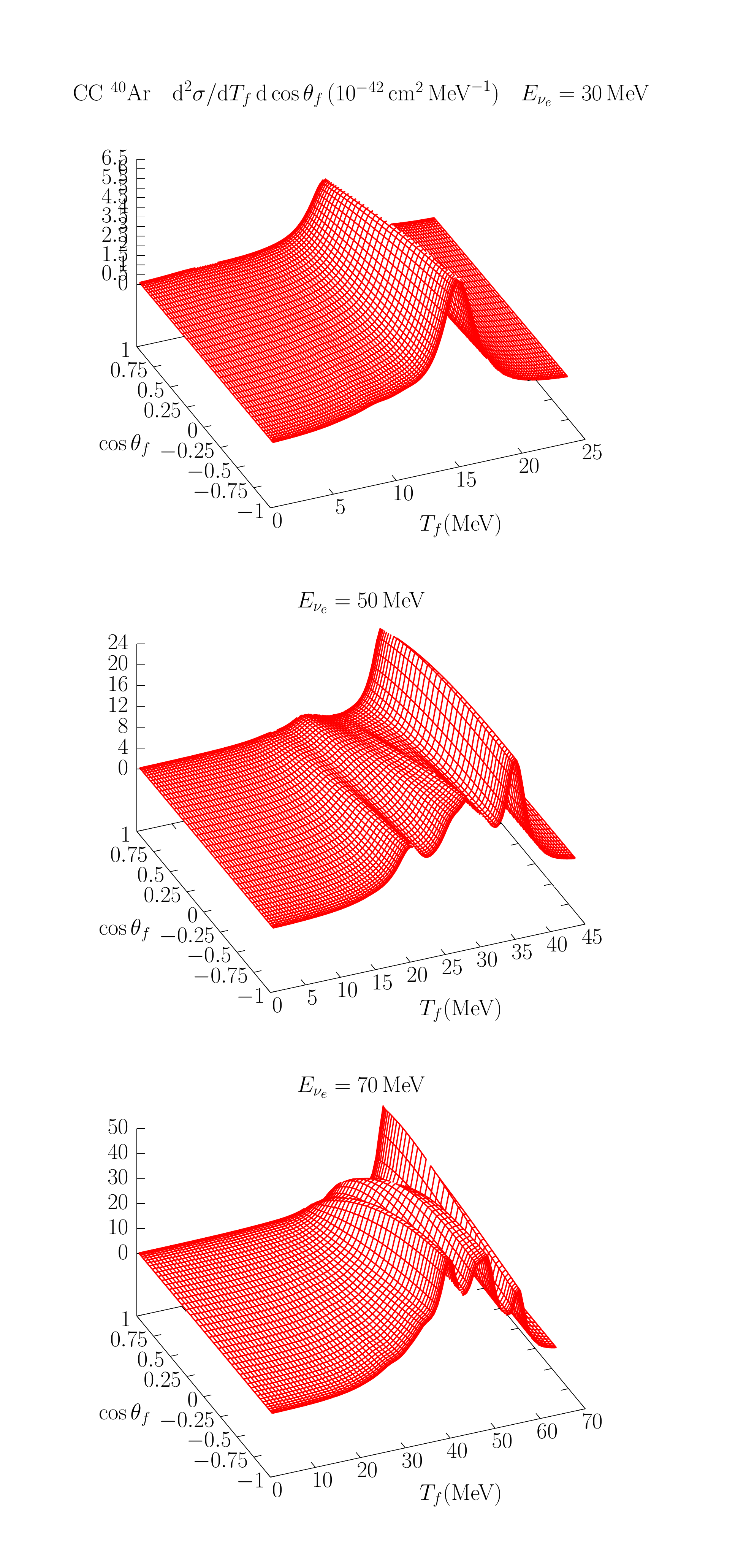}
   \caption{The double differential CC ($\nu_e$,$^{40}$Ar) cross section as a function of lepton scattering angle $\cos{\theta_f}$ and lepton kinetic energy $T_f$ for incoming neutrino energies 30, 50 and 70 MeV with contributions from different multipole moments. The cross section was folded with a Lorentzian of width 3 MeV to account for the finite width of nuclear excitations.}
   \label{fig:ddiffarcc}
\end{figure}

\begin{figure}
   \centering
   \includegraphics[width=0.95\columnwidth]{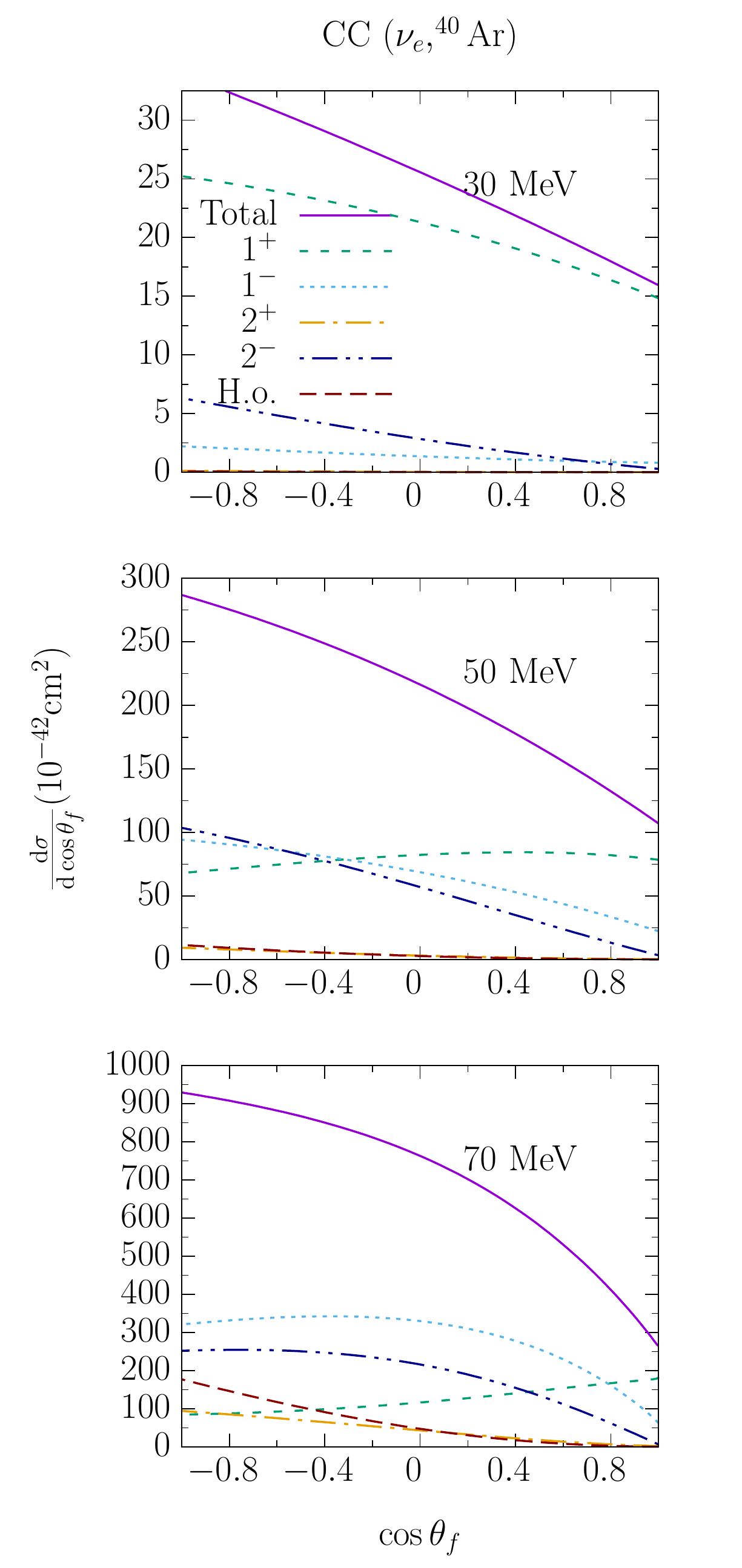}
   \caption{The single differential CC ($\nu_e$,$^{40}$Ar) cross section as a function of lepton scattering angle $\cos{\theta_f}$ for incoming neutrino energies 30, 50 and 70 MeV with contributions from different multipole moments. 'H.o.' contains the sum of all remaining higher--order multipole contributions.}
   \label{fig:monoarcc}
\end{figure}

\begin{figure}
   \centering
   \includegraphics[width=0.95\columnwidth]{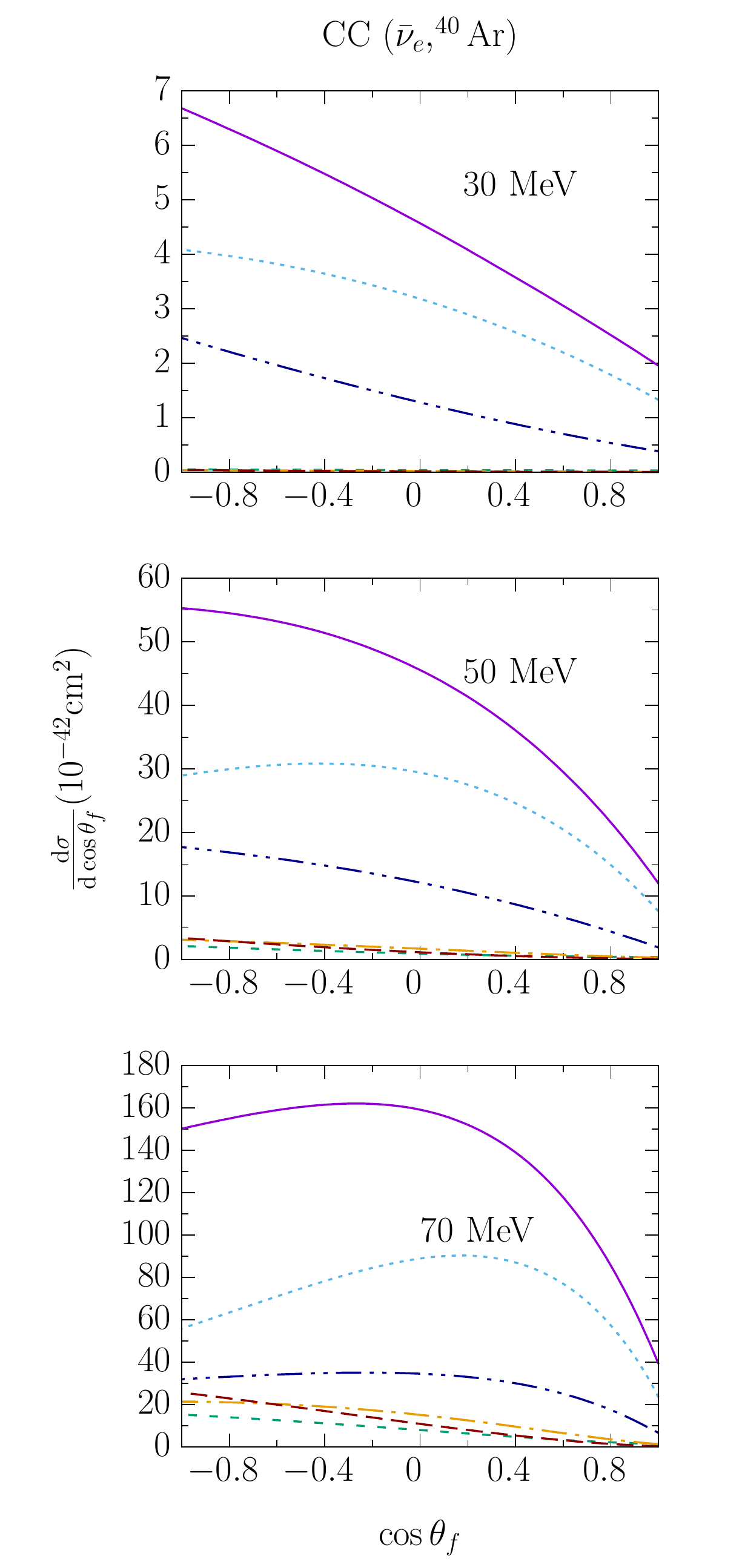}
   \caption{The differential CC ($\bar{\nu}_e$,$^{40}$Ar) cross section as a function of lepton scattering angle $\cos{\theta_f}$ for incoming neutrino energies 30, 50 and 70 MeV with contributions from different multipole moments. Same key as in Fig.~\ref{fig:monoarcc}.}
   \label{fig:monoarccanti}
\end{figure}

\begin{figure}
   \centering
   \includegraphics[width=0.95\columnwidth]{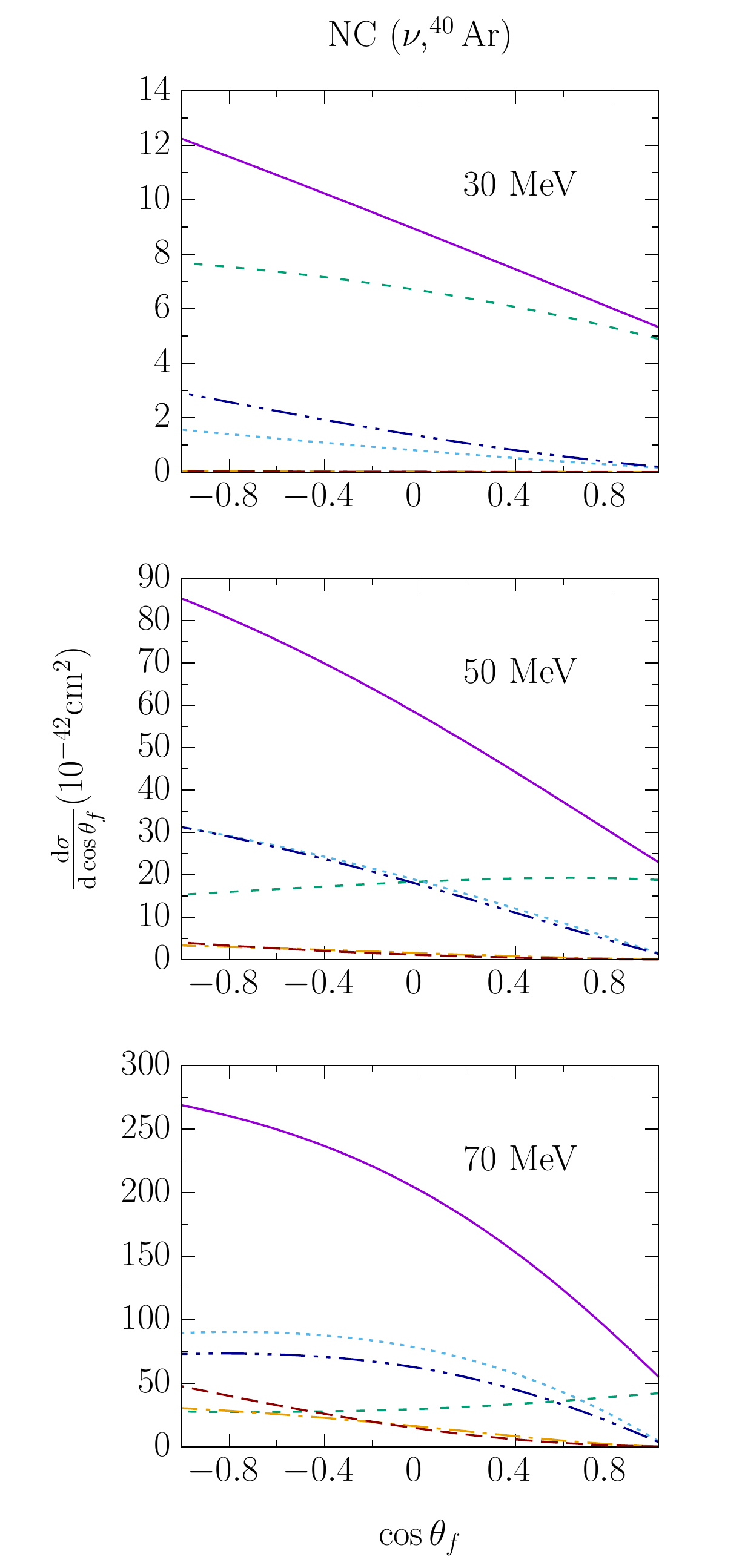}
   \caption{The differential NC ($\nu$,$^{40}$Ar) cross section as a function of lepton scattering angle $\cos{\theta_f}$ for incoming neutrino energies 30, 50 and 70 MeV with contributions from different multipole moments. Same key as in Fig.~\ref{fig:monoarcc}.}
   \label{fig:monoarnc}
\end{figure}

\begin{figure}
   \centering
   \includegraphics[width=0.95\columnwidth]{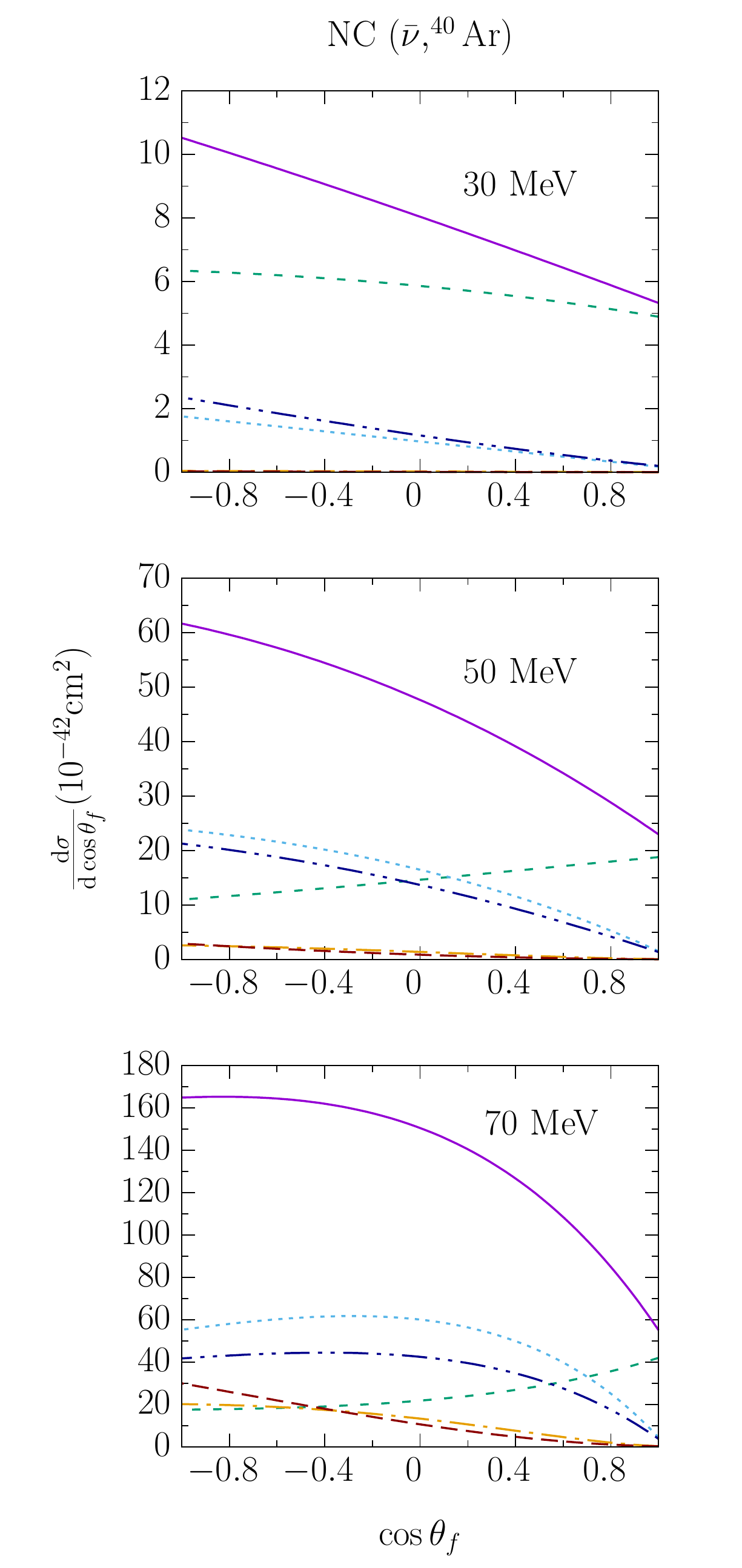}
   \caption{The differential NC ($\bar{\nu}$,$^{40}$Ar) cross section as a function of lepton scattering angle $\cos{\theta_f}$ for incoming neutrino energies 30, 50 and 70 MeV with contributions from different multipole moments. Same key as in Fig.~\ref{fig:monoarcc}.}
   \label{fig:monoarncanti}
\end{figure}

We begin our discussion by taking a look at CRPA differential cross section predictions for a variety of kinematical conditions applicable to low--energy scenarios. Shown in Figure~\ref{fig:ddiffarcc} are the double differential cross sections calculated for charged--current (CC) neutrinos scattering off $^{40}$Ar, as a function of the outgoing electron's kinetic energy $T_f$ and its scattering angle $\cos{\theta_f}$. As the energy increases, more resonance peaks show up as an increasing number of excitations becomes accessible. We can also integrate out the lepton's kinetic energy, and focus on the single differential cross sections as a function of the direction the outgoing lepton scattering angle $\cos{\theta_f}$ in \Cref{fig:monoarcc,fig:monoarccanti,fig:monoarnc,fig:monoarncanti}. These plots present results for CC and NC reactions, for both neutrinos and antineutrinos, for incoming energies of 30, 50 and 70 MeV. In doing so, we can now show the separated contributions coming from the individual multipole moments of the nuclear current. 

\begin{figure}
   \centering
   \includegraphics[width=0.95\columnwidth]{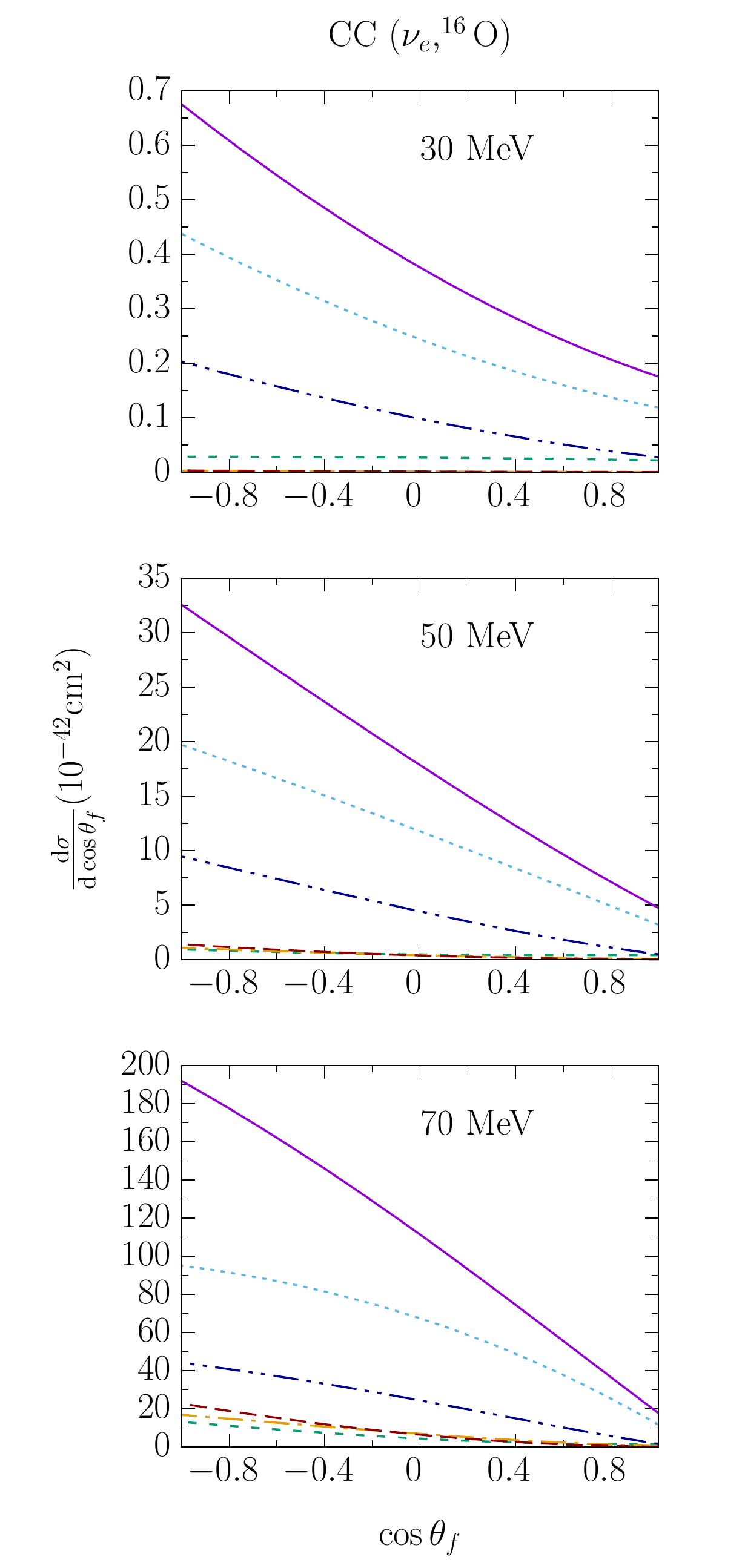}
   \caption{The single differential CC ($\nu_e$,$^{16}$O) cross section as a function of lepton scattering angle $\cos{\theta_f}$ for incoming neutrino energies 30, 50 and 70 MeV with contributions from different multipole moments. Same key as in Fig.~\ref{fig:monoarcc}.}
   \label{fig:monoocc}
\end{figure}

\begin{figure}
   \centering
   \includegraphics[width=0.95\columnwidth]{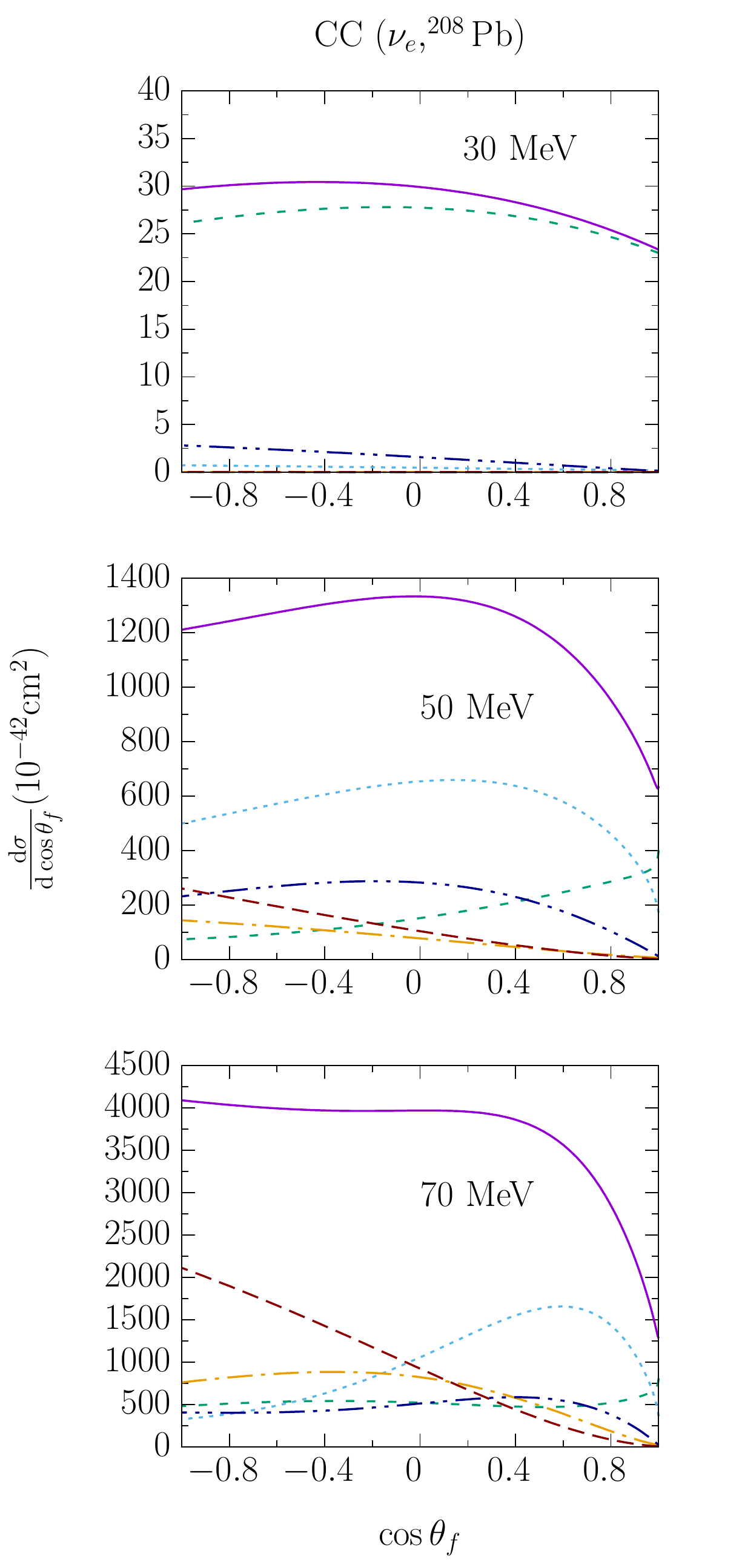}
   \caption{The single differential CC ($\nu_e$,$^{208}$Pb) cross section as a function of lepton scattering angle $\cos{\theta_f}$ for incoming neutrino energies 30, 50 and 70 MeV with contributions from different multipole moments. Same key as in Fig.~\ref{fig:monoarcc}.}
   \label{fig:monopbcc}
\end{figure}

While working in the AA is fair at 30 MeV in the case of neutrino--induced CC events, the forbidden transitions are needed to capture the full reaction strength for antineutrino--induced reactions, NC interactions and especially at higher energies for all channels. We mention at this stage that the $0^+$ and $0^-$ transitions contribute only minimally to the total reaction strength and are not shown separately, to improve the overal clarity of the plots. The Fermi transitions induced by these operators are not included in the CRPA results, as it only contains the continuous part of the excitation spectrum. Moreover, one can also appreciate the shape difference in the angular distribution between the $1^+$ contribution and the total differential cross section. We can also calculate these distributions for other nuclei. In comparing Figs.~\ref{fig:monoarcc}, \ref{fig:monoocc} and \ref{fig:monopbcc}, two things become clear concerning the A--dependence of the angular distributions. Firstly, we affirm the general expected trend that the number of multipoles needed for a satisfactory convergence of the differential cross section increases with the incoming energy of the neutrino and also with the increasing mass of the struck nucleus. A second observation we note is the qualitative way in which the angular distribution changes across energy and nuclear mass. Generally, the main feature is for these differential cross sections to be dominated by backward scattering. At the lowest energies, this is especially the case, but gradually less so as the energy increases. Similarly, the higher the mass of the struck nucleus, the less the differential cross section tends towards backwards scattering. Indeed, at 50 MeV, $^{16}$O and $^{40}$Ar paint quite a different picture from $^{208}$Pb, whichwhere cross sections peaks around $\cos{\theta_f} \approx 0$.

\begin{figure}
   \centering
   \includegraphics[width=0.95\columnwidth]{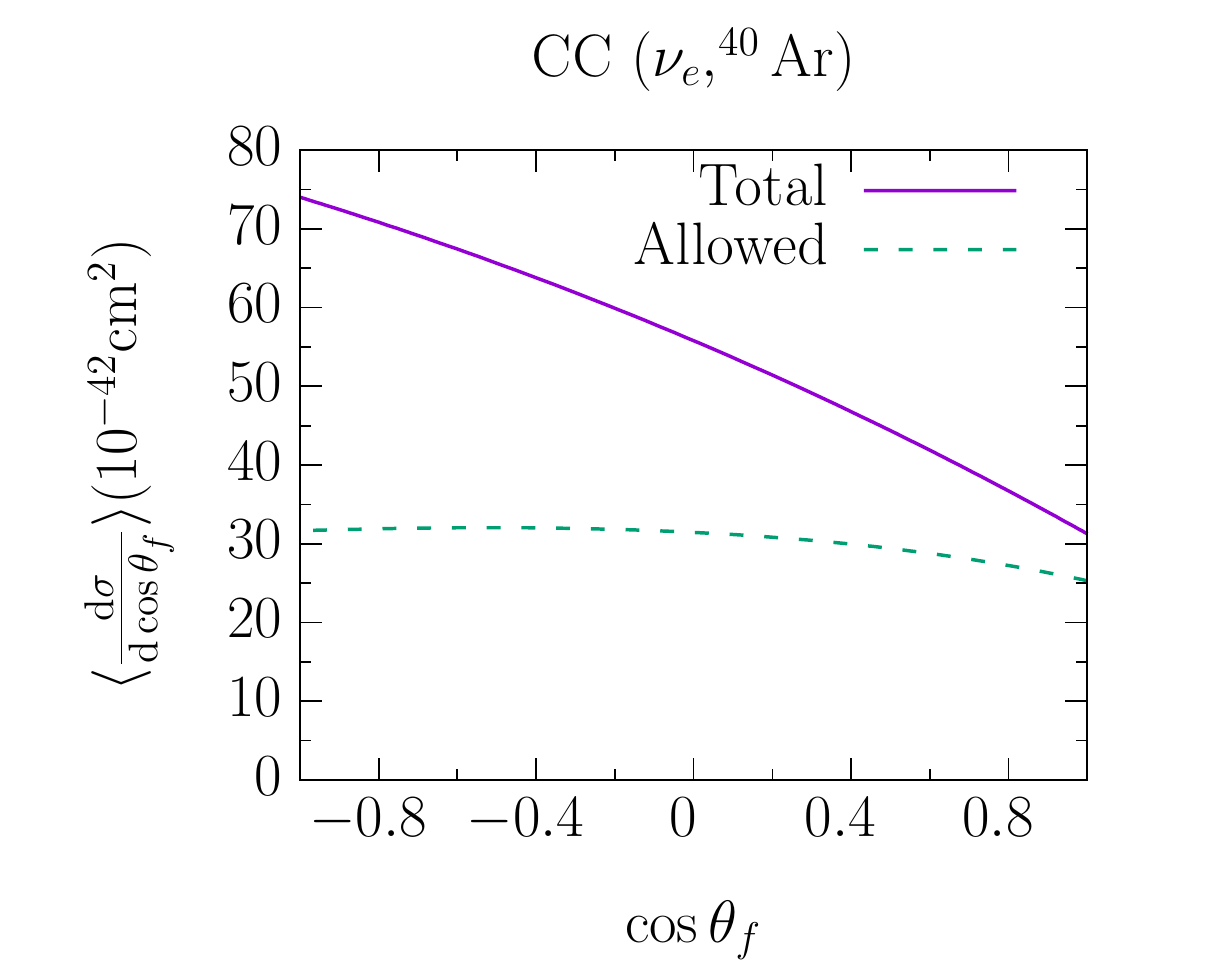}
   \caption{The differential CC ($\nu_e$,$^{40}$Ar) cross section as a function of lepton scattering angle $\cos{\theta_f}$ for a DAR neutrino energy spectrum, with the contribution of the allowed approximation separated.}
   \label{fig:dararcc}
\end{figure}

\begin{figure}
   \centering
   \includegraphics[width=0.95\columnwidth]{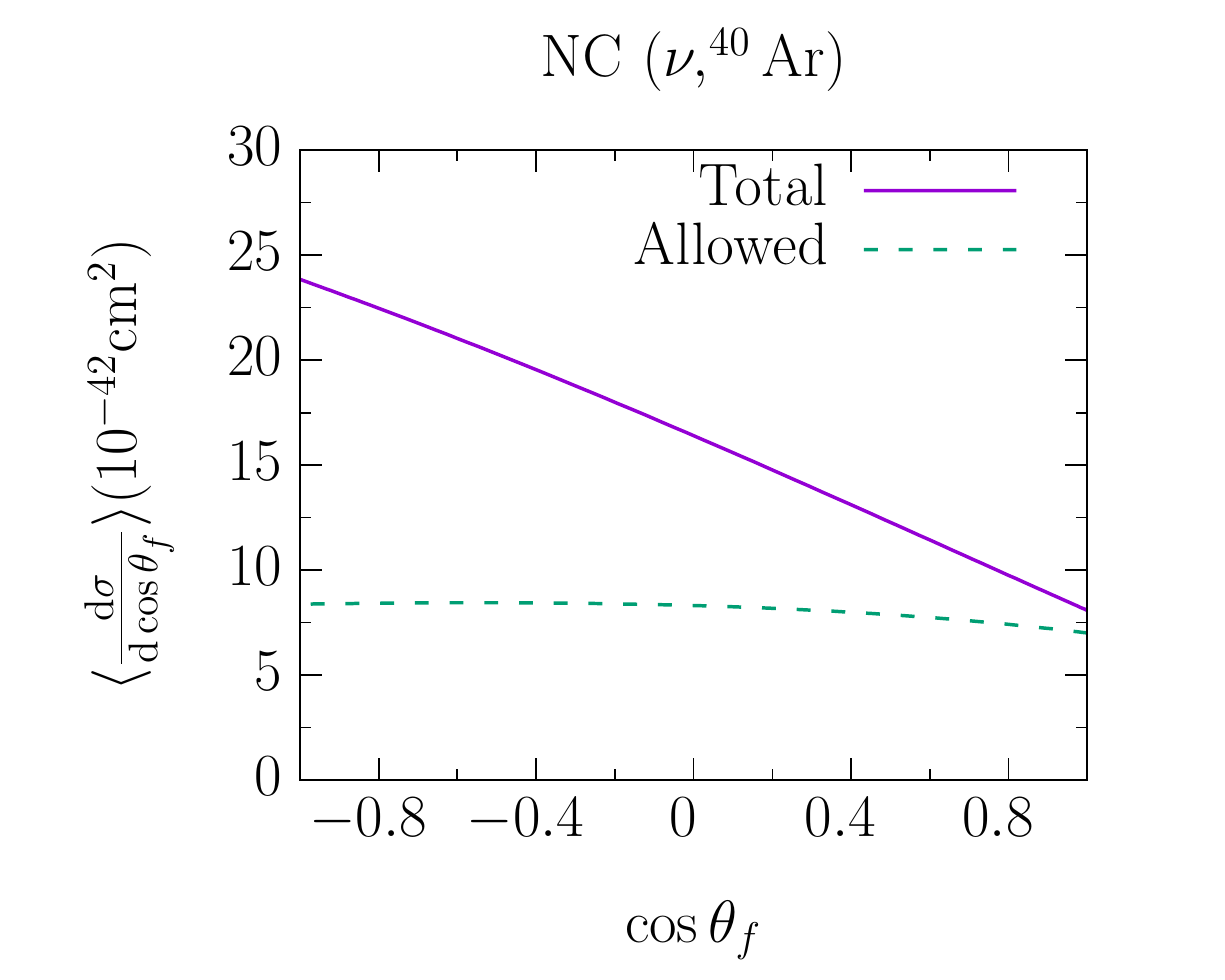}
   \caption{The differential NC ($\nu$,$^{40}$Ar) cross section as a function of lepton scattering angle $\cos{\theta_f}$ for a DAR neutrino energy spectrum, with the contribution of the allowed approximation separated.}
   \label{fig:dararnc}
\end{figure}

\begin{figure}[!hb]
   \centering
   \includegraphics[width=0.95\columnwidth]{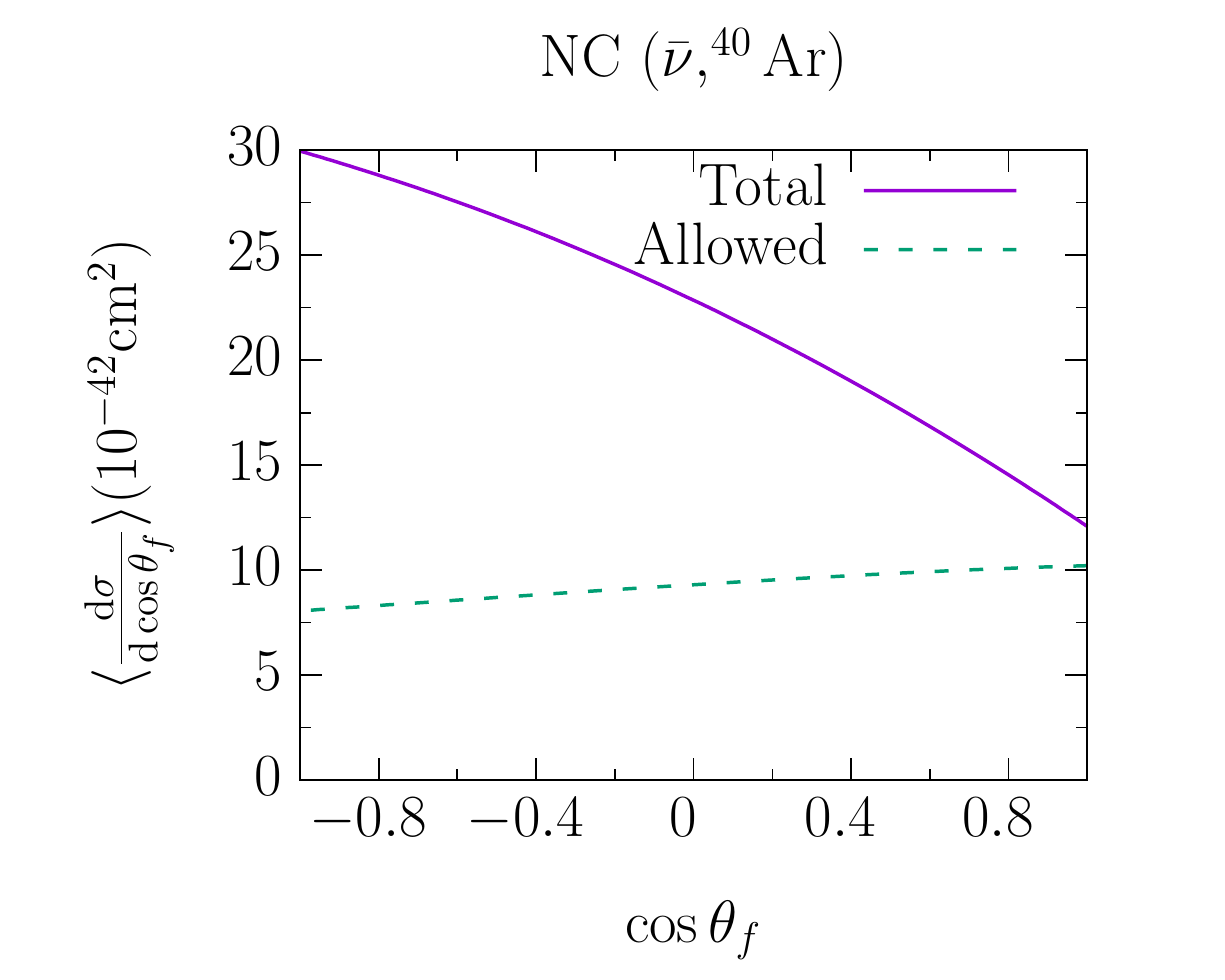}
   \caption{The differential NC ($\bar{\nu}$,$^{40}$Ar) cross section as a function of lepton scattering angle $\cos{\theta_f}$ for a DAR antineutrino energy spectrum, with the allowed approximation shown separately.}
   \label{fig:dararncanti}
\end{figure}

To get a better sense as to how big this shape difference has an effect in an experimental context, one can take a look at angular distributions folded with neutrino spectra, such as those yielded by pion decay--at--rest, or in the case of supernova neutrinos, a Fermi--Dirac distribution at various temperatures. We show the former for all possible scenarios (CC/NC, $\nu$/$\bar{\nu}$) except antineutrino--induced CC, since this is kinematically inaccessible. (The DAR spectrum only contains muon--antineutrinos, with not enough phase space available for an outgoing massive muon in the regimes discussed in this work.) The results are shown in \Cref{fig:dararcc,fig:dararnc,fig:dararncanti}. The AA predicts that leptons will be emitted from the reaction nearly isotropically. Taking forbidden transitions into account changes this picture, and shows that backward scattering is enhanced significantly, deviating strongly from the isotropic behavior of the AA contributions.

\begin{figure}[!ht]
   \centering
   \includegraphics[width=0.95\columnwidth]{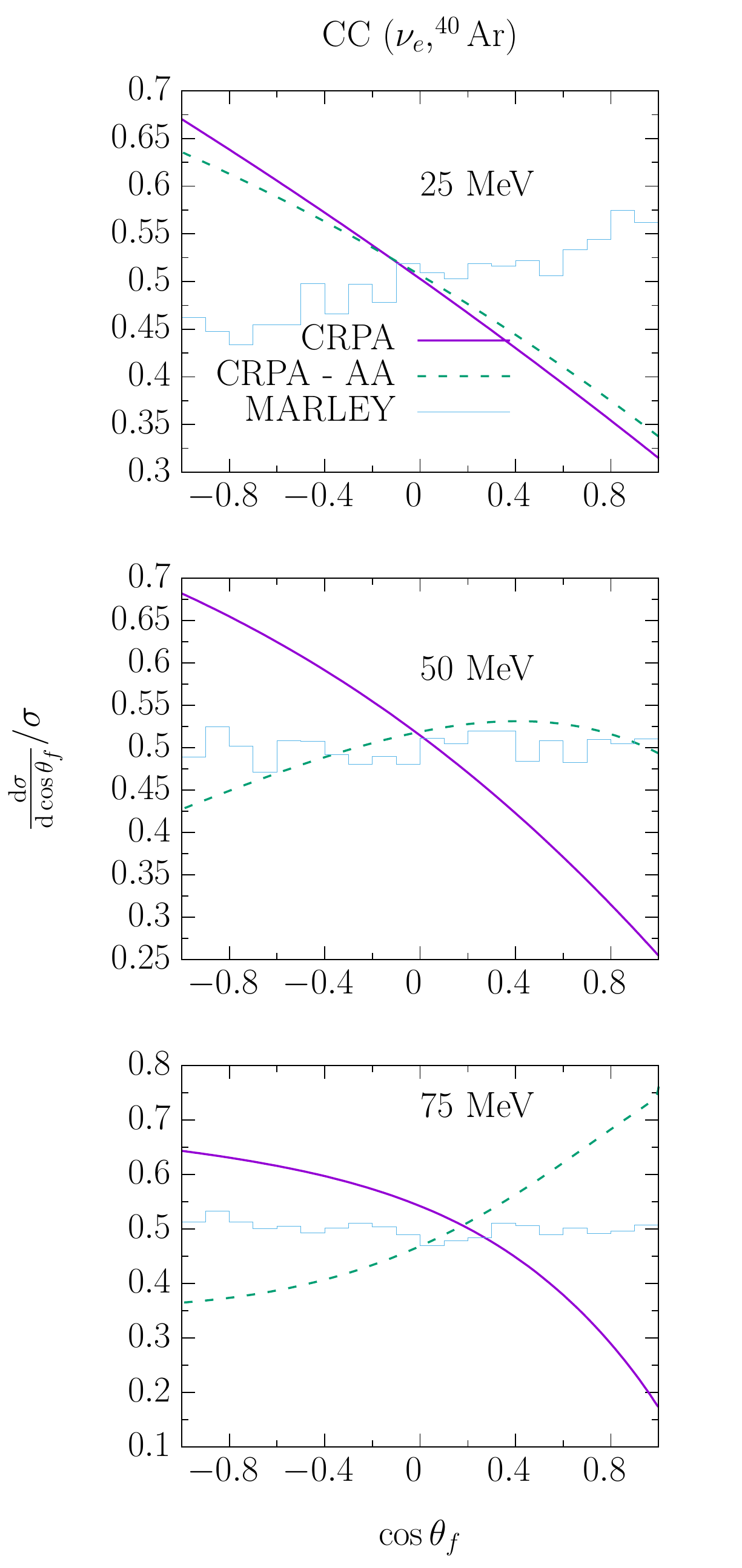}
   \caption{Normalized angular distributions for CC ($\nu_e$,$^{40}$Ar)  reactions as a function of lepton scattering angle $\cos{\theta_f}$ for mono--energetic neutrinos, with the MARLEY results in the histogram.}
   \label{fig:monoarcccomp}
\end{figure}

\begin{figure}[!ht]
   \centering
   \includegraphics[width=0.95\columnwidth]{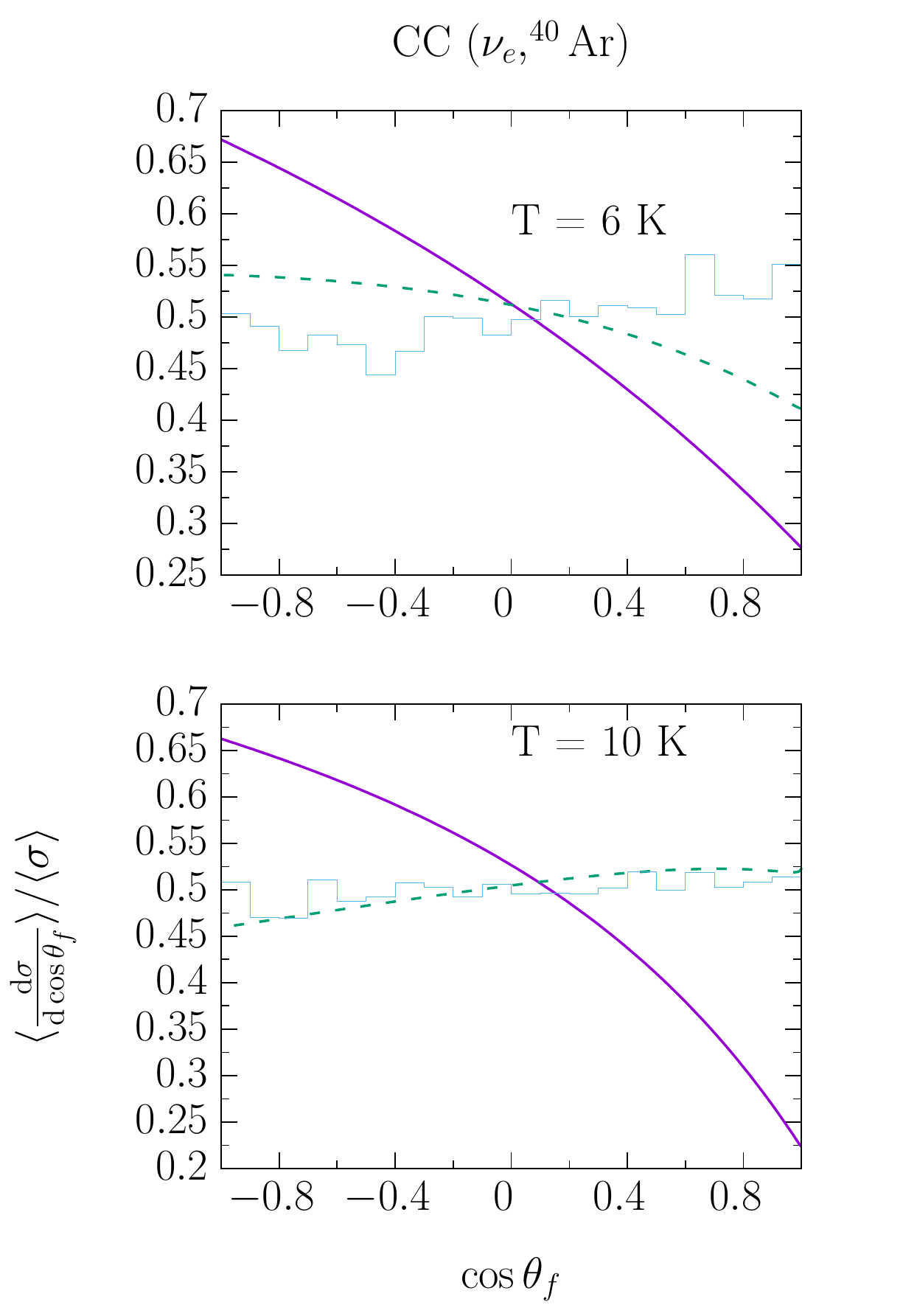}
   \caption{Normalized angular distributions for CC ($\nu_e$,$^{40}$Ar)  reactions as a function of lepton scattering angle $\cos{\theta_f}$ for a Fermi--Dirac distribution at various temperature parameters, with the MARLEY results in the histogram. Key identical to Fig.~\ref{fig:monoarcccomp}.}
   \label{fig:fermarcccomp}
\end{figure}

\begin{figure}
   \centering
   \includegraphics[width=0.95\columnwidth]{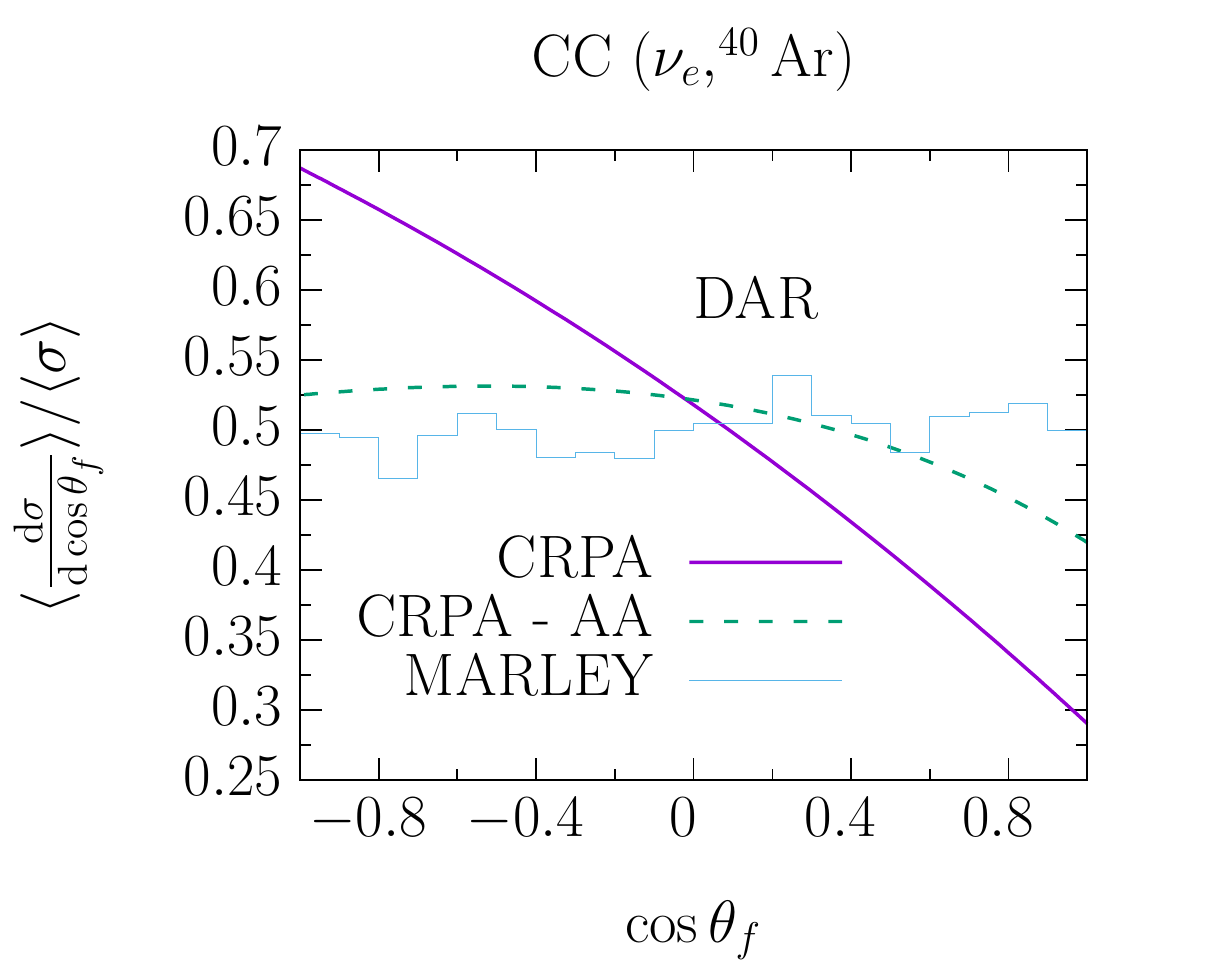}
   \caption{Normalized angular distributions for CC ($\nu_e$,$^{40}$Ar)  reactions as a function of lepton scattering angle $\cos{\theta_f}$ for a pion decay--at--rest (DAR) neutrino spectrum.}
   \label{fig:dararcccomp}
\end{figure}

Finally, we will compare these results with those of MARLEY (Modeling of Argon Reaction Low--energy Yields), a Monte Carlo (MC) event generator aimed at simulating neutrino--induced CC reactions on Argon for low--energy neutrinos~\cite{Gardiner2017}. This comparison is done in Figs.~\ref{fig:monoarcccomp},~\ref{fig:fermarcccomp} and \ref{fig:dararcccomp}.

These figures show that working in the allowed approximation, again, yields distributions that are approximately isotropical. This, too, is predicted by the MARLEY generator. The only exception seems to be at 25 MeV, where the MC distribution differs quite significantly from the CRPA predictions. As discrete excitations important at lower energies are not included in the CRPA spectrum, this is not surprising. At higher energies, the difference between the AA and the full CRPA model shows that including forbidden transitions causes the angular distribution to skew further towards backwards scattering. These forbidden transitions are not present in MARLEY. Since they clearly affect the kinematics of the final lepton, this can introduce biases in the experimental analyses, and should be taken into account. 
\begin{figure}
   \centering
   \includegraphics[width=0.95\columnwidth]{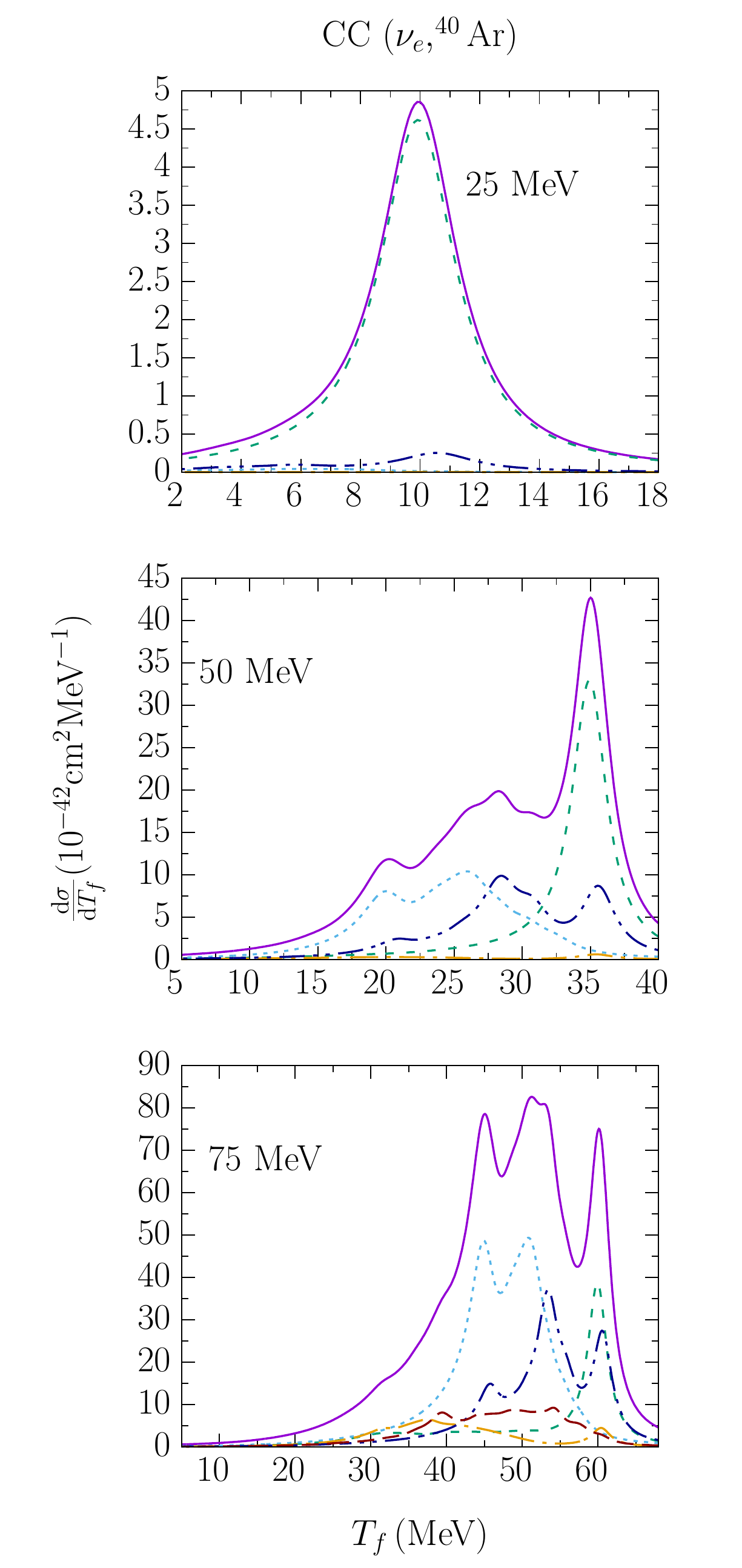}
   \caption{Differential cross sections for CC ($\nu_e$,$^{40}$Ar)  reactions as a function of lepton kinetic energy for several incoming neutrino energies. Same key as in Fig.~\ref{fig:monoarcc}.}
   \label{fig:monoarcctf}
\end{figure} 
Following on from this discussion on angular distributions, one can also look at the distributions in terms of the final lepton's kinetic energy $T_f$. We do so for CC neutrino--argon scattering at several energies in Fig.~\ref{fig:monoarcctf}, where the differential cross section as a function of $T_f$ is plotted with the individual multipole contributions separated, similar to Fig.~\ref{fig:monoarcc}. The conclusions are of course the same, except that the shape differences in the distributions as a consequence of forbidden transitions are even more remarkable here. 
\begin{figure}
   \centering
   \includegraphics[width=0.95\columnwidth]{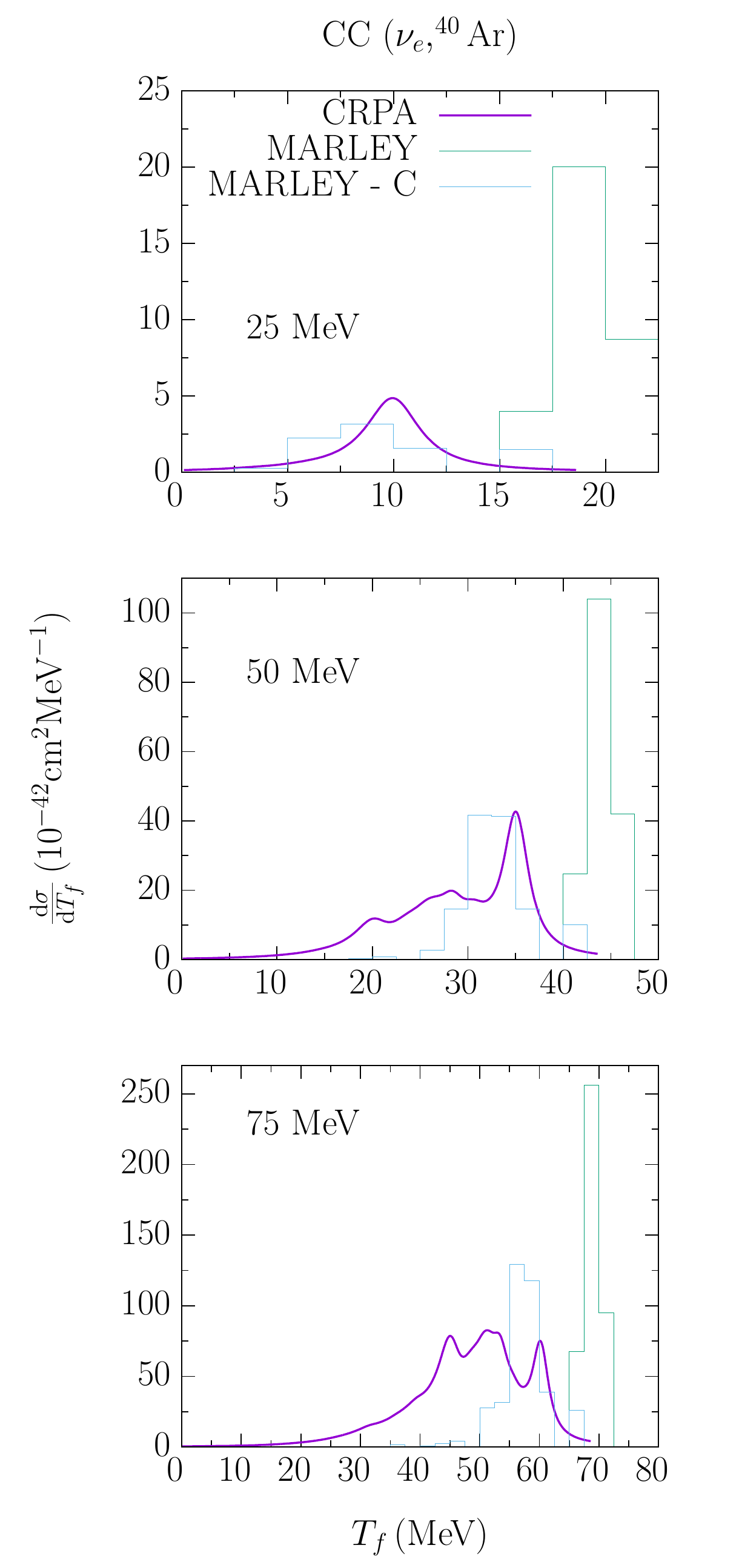}
   \caption{Differential cross sections for CC ($\nu_e$,$^{40}$Ar)  reactions as a function of lepton kinetic energy for several incoming neutrino energies, compared to MARLEY, with continuous excitations (MARLEY--C) shown separately.}
   \label{fig:monoarcctfmarley}
\end{figure}
We can now compare these to MARLEY predictions, where we can make an explicit distinction between the discrete and continuous excitations. This is shown in Fig.~\ref{fig:monoarcctfmarley}. In this figure it is shown that, while the CRPA model does not model the discrete excitations, there is a shift towards lower kinetic energies for the outgoing lepton present in the CRPA model. As is shown in Fig.~\ref{fig:monoarcctf}, caused almost exclusively by the contribution of forbidden transitions. This discrepancy gets worse as the energy increases. 
\begin{figure}[!ht]
   \centering
   \includegraphics[width=0.95\columnwidth]{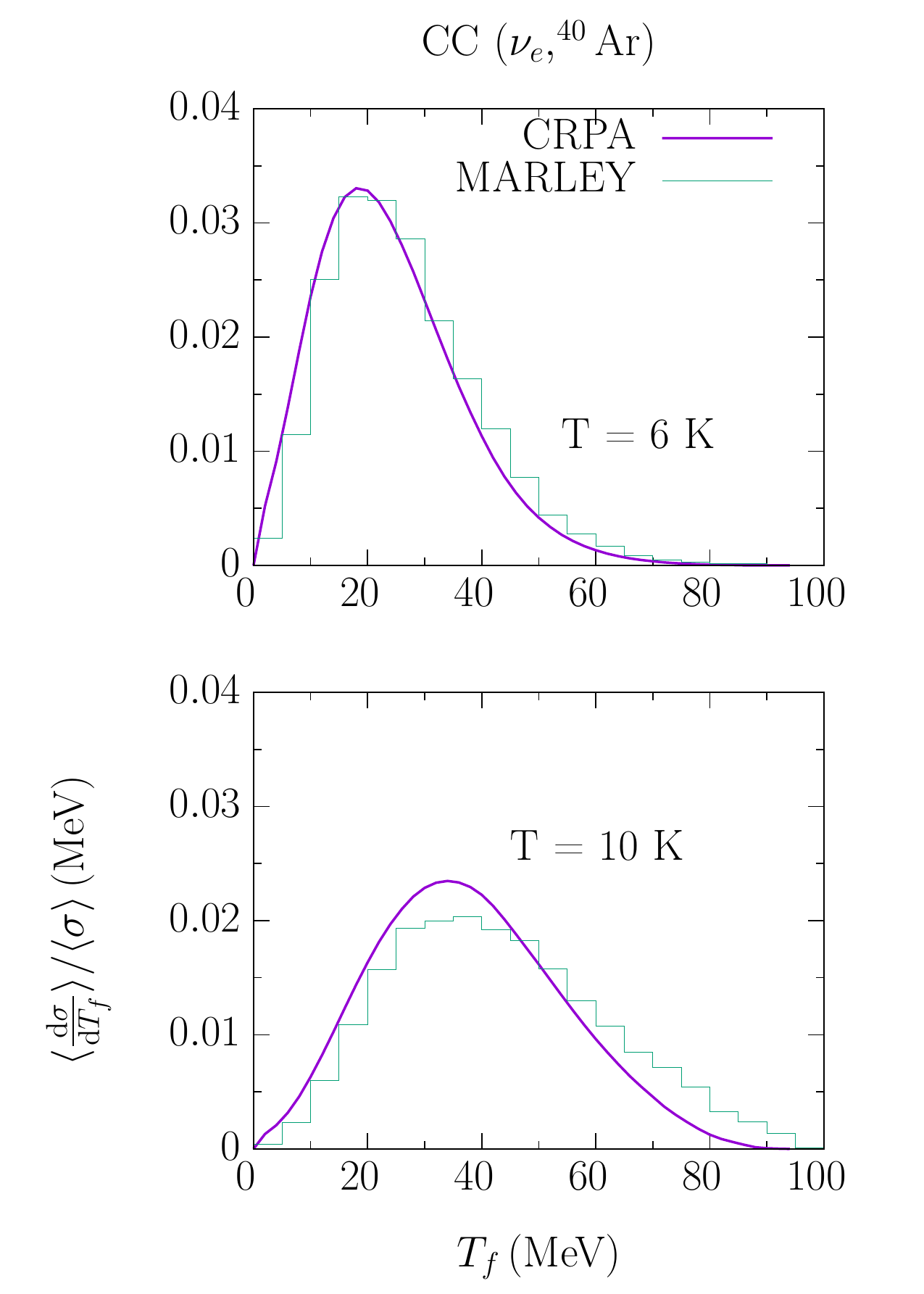}
   \caption{Normalized kinetic energy distributions for CC ($\nu_e$,$^{40}$Ar) reactions as a function of lepton kinetic energy $T_f$ for a Fermi--Dirac distribution at various temperature parameters.}
   \label{fig:fermarcccomptf}
\end{figure} 
To get a sense of the effect this has on the signals in experimental situation, we show the normalized strength distribution for two Fermi--Dirac neutrino spectra, for two temperatures 6 K and 10 K in Fig.~\ref{fig:fermarcccomptf}. While the CRPA model and the MARLEY results match favorably for the low--temperature spectrum, the higher one will start to show the effect of the previously mentioned discrepancy of having more low $T_f$ events in the CRPA (see Fig.~\ref{fig:monoarcctfmarley}) due to forbidden transitions. Furthermore, contrary to our approach MARLEY extrapolates the matrix elements for the discrete low--energy transitions to higher energies without accounting for the q--dependence. This introduces a bias in e.g. energy reconstruction in experimental analyses, and should therefore be included.

\section{Summary}\label{sec:summ}

With Argon an important future target for low--energy neutrinos, we have presented differential cross section results for neutrino scattering, with a focus on $^{40}$Ar nuclei. The CRPA approach used here allows us to include the effects of long--range correlations and collectivity in the nuclear response. In this work we have focused on the angular distributions of the final lepton created in this process. In doing so, we found that while forbidden transitions not only contribute non--trivially to the overal reaction strength, they also cause a re-shaping of the outgoing lepton's angular and kinetic energy distributions. This is demonstrated for several neutrino spectra, and compared with the MARLEY MC generator, which lacks forbidden transition modeling. Incorporating these could greatly improve on the quality of predictions and analyses.

 \begin{acknowledgments}
   This work was supported by the Research Foundation Flanders (FWO--Flanders). The computational resources (Stevin Supercomputer Infrastructure) and services used in this work were provided by the VSC (Flemish Supercomputer Center), funded by Ghent University, FWO and the Flemish Government -- Department EWI.
 \end{acknowledgments}

 \bibliography{biblio}
 \end{document}